# Algorithms for Estimating Information Distance with Application to Bioinformatics and Linguistics

Alexei Kaltchenko

*Department of Physics and Computing*
*Wilfrid Laurier University*
*Waterloo, Ontario N2L3C5, Canada*
akaltche@wlu.ca

**Abstract**
*After reviewing unnormalized and normalized information distances based on incomputable notions of Kolmogorov complexity, we discuss how Kolmogorov complexity can be approximated by data compression algorithms. We argue that optimal algorithms for data compression with side information can be successfully used to approximate the normalized distance. Next, we discuss an alternative information distance, which is based on relative entropy rate (also known as Kullback-Leibler divergence), and compression-based algorithms for its estimation. Based on available biological and linguistic data, we arrive to unexpected conclusion that in Bioinformatics and Computational Linguistics this alternative distance is more relevant and important than the ones based on Kolmogorov complexity.*

## 1. INFORMATION DISTANCES BASED ON KOLMOGOROV COMPLEXITY

Suppose, for a positive integer $n$, we have two strings of characters $x \triangleq (x_1, x_2, x_3, \ldots, x_n)$ and $y \triangleq (y_1, y_2, y_3, \ldots, y_n)$ that describe two similar objects (such as DNA sequences, texts, pictures, etc). The strings are assumed to be drawn from the same alphabet $\mathcal{A}$. To quantify similarity between the objects, one needs a notion of information distance between two individual objects (strings). Bennett et al introduced[3] a distance function $E_1(x,y)$ defined by
$$E_1(x, y) \triangleq \max \{K(x \mid y), K(y \mid x)\},$$
where $K(x|y)$ is the conditional Kolmogorov complexity[8] of string $x$ relative to string $y$, defined as the length of a shortest binary program to compute $x$ if $y$ is furnished as an auxiliary input to the computation. They shown that the distance $E_1(x,y)$ is a universal *metric*. Formally, a distance function $d$ with nonnegative real values, defined on the Cartesian product $X \times X$ of a set $X$, is called a metric on $X$ if for every $x, y, z \in X$:

- $d(x,y) = 0$ iff $x = y$   (identity axiom)
- $d(x,y) + d(y,z) \geq d(x,z)$   (triangle inequality)
- $d(x,y) = d(y,x)$   (symmetry axiom)

The universality implies that if two objects are similar in some computable metric, then they are at least that similar in $E_1(x,y)$ sense.

A distance function is called *normalized* if it takes values in [0;1]. Thus, distance $E_1(x,y)$ is clearly unnormalized. Li at al argued[7] that in Bioinformatics an unnormalized distance may not be a proper evolutionary distance measure. It would put two long and complex sequences that differ only by a tiny fraction of the total information as dissimilar as two short sequences that differ by the same absolute amount and are completely random with respect to one another. They proposed a normalized information distance $E_2(x,y)$ defined by

$$E_2(x, y) \triangleq \frac{\max \{K(x \mid y), K(y \mid x)\}}{\max \{K(x), K(y)\}} \qquad (1)$$

and proved that it is a universal metric, too.

## 2. ESTIMATION OF DISTANCE $E_2(x,y)$ VIA COMPRESSION.

Since the information distance $E_2(.,.)$ is based on noncomputable notions of Kolmogorov complexities, we need to approximate the latter by computable means. It is well known that the Kolmogorov complexity and compressibility of strings are closely related[6],[8]. So we need data compression algorithms suitable for approximating Kolmogorov complexities.

To express function $E_2(x, y)$ via information-theoretic measures relevant to data compression, we can write distance $E_2(x, y)$ as

$$E_2(x, y) = \frac{\max\{\frac{1}{n}K(x|y), \frac{1}{n}K(y|x)\}}{\max\{\frac{1}{n}K(x), \frac{1}{n}K(y)\}} \quad (2)$$

Through the rest of this paper we assume[1] that strings $x$ and $y$ are generated by finite-order, stationary Markov sources $X$ and $Y$, respectively, and this source pair jointly forms a finite-order, stationary Markov source, too. Then, from Information Theory, we have the following almost sure convergence:

$$\lim_{n \to \infty} \tfrac{1}{n} K(x) = H(X), \text{ a.s.}$$
$$\lim_{n \to \infty} \tfrac{1}{n} K(x|y) = H(X|Y), \text{ a.s.} \quad (3)$$

Thus, we have

$$\lim_{n \to \infty} E_2(x, y) = \frac{\max\{H(X|Y), H(Y|X)\}}{\max\{H(X), H(Y)\}}, \text{ a.s.} \quad (4)$$

Consequently, the right-hand side of (4) can be used as a good approximation of $E_2(x, y)$ for sufficiently large $n$. Thus, we are interested in data compression algorithms capable of estimating entropy rate $H(\bullet)$ and conditional entropy rate $H(\bullet|\bullet)$. In view of the information-theoretic identity

$$H(X|Y) = H(X,Y) - H(Y), \quad (5)$$

where $H(X,Y)$ denotes joint entropy rate, we can also estimate conditional entropy rate $H(\bullet|\bullet)$ indirectly, via estimation of joint entropy rate $H(\bullet, \bullet)$ and (unconditional) entropy rate $H(\bullet)$. In the following three subsections, we will discuss data compression algorithms for estimation $H(\bullet)$, $H(\bullet, \bullet)$, and $H(\bullet|\bullet)$, respectively.

## 2.1 Estimation of Entropy Rate

For any data compression algorithm, we define its compression rate by the ratio $|b_x|/|x|$, where $b_x$ is the binary codeword produced by the algorithm for input string $x$, and $|\cdot|$ denotes string length. Then, $H(X)$ can be approximated by the compression rate of an optimal lossless data compression algorithm as shown in Figure 1. From Information Theory, the optimality implies that, as the length of $x$ grows, $|b_x|/|x|$ converges to $H(X)$.

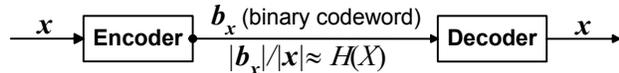

**Figure 1: Optimal compression of strings**

---

[1] This is a common assumption for compression analysis.

In [7], Li et al used GenCompress[5], an efficient algorithm for DNA sequence compression, for estimating Kolmogorov complexities. Based on good compression performance of GenCompress, we can reasonably assume that the algorithm is optimal or near-optimal, and, thus, the rate $|b_x|/|x|$, provides a good estimate for $H(X) \approx \frac{1}{|x|} K(x)$. As for estimating joint Kolmogorov complexity $K(x, y)$, Li et al used the concatenated sequence $x + y \triangleq (x_1, x_2, x_3, \ldots, x_n, y_1, y_2, y_3, \ldots, y_n)$ as the input to GenCompress as shown in Figure 2. They heuristically assumed that the size of the compressed output would approximate $K(x, y)$, which would imply that the ratio $|b_{x+y}|/|x|$ would approximate $H(X,Y)$.

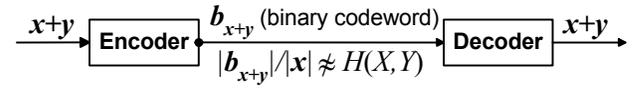

**Figure 2: Non-optimal compression of string pairs.**

However, from Information Theory, this assumption is generally not correct because encoding the concatenated string $x+y$ as a single string from the same alphabet $\mathcal{A}$ does not properly utilize the correlation between $x$ and $y$. To properly utilize the correlation, a compression algorithm must encode the $(x,y)$ pair as the string $s \triangleq \left(\binom{x_1}{y_1}, \binom{x_2}{y_2}, \binom{x_3}{y_3}, \ldots, \binom{x_n}{y_n}\right)$ of supersymbols from the alphabet $\mathcal{A} \times \mathcal{A}$

## 2.2 Estimation of Joint Entropy Rate

$H(X,Y)$ can be approximated by the compression rate of an optimal lossless data compression algorithm as shown in Figure 3. From Information Theory, the optimality implies that, as the length of $x$ and $y$ grows, $|b_{x,y}|/|x|$ converges to $H(X,Y)$.

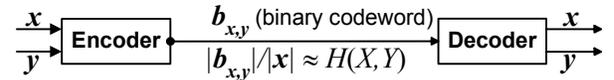

**Figure 3: Optimal compression of string pairs.**

As we discussed in the previous section, for its optimal compression, a string pair $(x,y)$ must be encoded as a string $s \triangleq \left(\binom{x_1}{y_1}, \binom{x_2}{y_2}, \binom{x_3}{y_3}, \ldots, \binom{x_n}{y_n}\right)$ of supersymbols from the alphabet $\mathcal{A} \times \mathcal{A}$ Moreover, we can simply compress string $s$ by any optimal algorithm for string compression depicted in Figure 1. Then, from Information Theory, we have: $|b_s|/|s| \approx H(X,Y)$

## 2.3 Estimation of Conditional Entropy Rate via Compression with Side Information

$H(X|Y)$ can be approximated by the compression rate of an optimal lossless data compression algorithm with side information as shown in Figure 4. From Information Theory, the optimality implies that, as the length of $x$ and $y$ grows, $|b_{x|y}|/|x|$ converges to $H(X|Y)$.

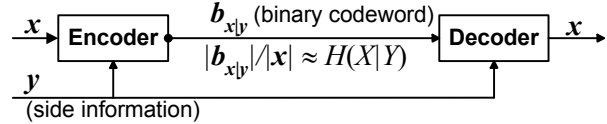

**Figure 4: Optimal compression of string $x$ with string $y$ used as side information.**

An optimal (in the above convergence sense) algorithm for universal lossless data compression with side information was proposed and analyzed in [9]. It is important to note that, as in the case with estimation of joint entropy rate, the algorithm encodes a string pair $(x,y)$ as a string of supersymbols from the alphabet $\mathcal{A} \times \mathcal{A}$

## 2.3 Estimation of Conditional Entropy Rate: Direct vs. Indirect

While $H(X,Y)$ is never less than $H(Y)$, the estimate of $H(X,Y)$ can be less than the estimate of $H(Y)$ for some string pairs, especially for those not sufficiently long. Thus, we can get a negative estimate of conditional entropy rate in the indirect estimation via identity (5). This may result in a negative information distance, which in turn would adversely affect building distance-based phylogeny trees in Bioinformatics and Computational Linguistics. Clearly, the direct estimation of conditional entropy $H(X|Y)$ via compression with side information will never yield a negative value (regardless of the estimation accuracy).

## 3. INFORMATION DISTANCE BASED ON RELATIVE ENTROPY RATE

Let $p_x$ and $q_z$ be probability measures for sources $X$ and $Z$, respectively. The relative entropy rate $D(Z\|X)$ (also known as Kullback-Leibler divergence) is defined by

$$D(Z\|X) \triangleq \limsup_{n \to \infty} \frac{1}{n} \sum_x q_z(x) \log \frac{q_z(x)}{p_x(x)}.$$

$D(Z\|X)$ is a nonnegative continuous function (functional) and equals to zero if and only if $p_x$ and $q_z$ coincide. Thus, $D(Z\|X)$ can be naturally viewed as a distance between the measures $p_x$ and $q_z$. However, $D(\cdot\|\cdot)$ is not a metric because it generally is neither symmetric, nor satisfies the triangle inequality.

It is not difficult to see that we can have $D(Z\|X)$ equal zero while the conditional entropy rate $H(Z|X)$ is large and vice versa. Thus, an information distance based on relative entropy rate may be a complement or even an alternative to an information distance based on Kolmogorov complexity.

## 3.1 Estimation of Relative Entropy via Data Compression

In data compression, the relative entropy $D(Z\|X)$ has the following interpretation of compression non-optimality. Loosely speaking, if we have a compression code (mapping), which optimally compresses string $x$ and use this code to (non-optimally) compress string $z$, then the compression rate will be $H(Z) + D(Z\|X)$. On the other hand, a compression code, which is optimal for string $z$, will compress it at the rate of $H(Z)$. Thus, by subtracting the latter from the former, we will get an estimate of $D(Z\|X)$. Compression-based algorithms for estimation of $D(\cdot\|\cdot)$ were proposed and analyzed in [4] and [10]. Work [4] also includes many interesting simulation results. Yet another compression-based algorithm for $D(\cdot\|\cdot)$ estimation was introduced in [2]. It was however purely heuristic and there was no claim that the algorithm converges to $D(\cdot\|\cdot)$.

**Conjecture:**
If we consider non-optimal compression of a concatenated string $x+y$ by GenCompress discussed in Section 2.1, then, for sufficiently large strings,
$$|b_{x+y}|/|x| \approx H(X) + H(Y) + D(Y\|X).$$

Thus, instead of $K(x|y)$, Li et al[7] could have actually estimated
$$|b_{x+y}|/|x| - |b_y|/|y| \approx H(X) + D(Y\|X).$$

They used this estimates to build (1) the mammalian DNA evolutionary tree and (2) the language classification tree, where each language was represented by a file of "The Universal Declaration of Human Rights" in that language. Since the Declaration was originally created in English and then "losslessly" translated into the other languages, the value $H(X)$ for every language is approximately the same[1]. Thus, instead of $E_2(x,y)$, Li et al could have actually estimated
$$\max\{D(Y\|X), D(X\|Y)\} + const.$$

Our conjecture is directly supported by the computational results in [2],[4], where the language classification tree was built with symmetrized information distance function based on relative entropy. The conjecture is also indirectly supported by the following fact. The mammalian DNA evolutionary tree, which was built in [7] based on presumably incorrect estimates of $K(\bullet|\bullet)$, is nevertheless correct. Thus, in Bioinformatics and Linguistics, an information distance based on relative entropy appears to be more meaningful than a distance based on Kolmogorov complexity.

## 6. CONCLUSIONS

After reviewing unnormalized and normalized information distances based on notions of Kolmogorov complexity, we propose compression algorithms for Kolmogorov complexity estimation. In particular, we suggest a method for direct estimation of conditional Kolmogorov complexity, which always yields a nonnegative value. We point out the limitations of the approach for estimating joint Kolmogorov complexity presented in [7].

We also discuss an alternative information distance based on relative entropy rate (also known as Kullback-Leibler divergence) and compression-based algorithms for its estimation.

Based on the computational results for the DNA evolutionary tree[7] and for the language classification tree[2],[4],[7], we conjecture that in Bioinformatics and Linguistics, an information distance based on relative entropy rate is more relevant and important than distances based on Kolmogorov complexity.

## Acknowledgements

The author thanks Ming Li for valuable comments on this work.